\documentclass[submission, LectureNotes]{SciPost}
\usepackage{amsmath}
\usepackage{gensymb}

\begin{document}

\begin{center}{\Large \textbf{ A new and coherent interstitial-ice model for pure water part II: explaining the conflicting Hall data }}\end{center}

\begin{center}
J. De Poorter\textsuperscript{1}
\end{center}

\begin{center}
{\bf 1} Horatio vzw - Koningin Maria Hendrikaplein 64d, Ghent, 9000, Belgium
\\
* john.zarat@gmail.com
\end{center}

\begin{center}
\today
\end{center}


\section*{Abstract}
{\bf
In part I of this paper the electric behaviour of pure water is described by an interstitial-ice model, the so-called Protonic-Semiconductor Interstitial-Ice or PSII model. Liquid water consists of an intact ice-like lattice with a significant percentage of both vacant lattice positions and water molecules filling the interstitial sites of the open ice-like lattice. It is shown that not the Grotthuss mechanism is dominant in water but a thermally-induced hopping mechanism of the H$^+$ and OH$^-$ ions linked to vacancies. In part II this hopping mechanism is further explored and confronted to the small Hall mobilities of the ions, still unexplained with mainstream models. Two types of complexes are found. The first type is composed of either charged vacancies with the opposite charge of the ion (H$^+$VL and OH$^-$VD) or neutral vacancies (H$^+$V and OH$^-$V). These complexes are responsible for the major part of the mobility of the ions. However, their movement is changing the electric polarisation density of the lattice structure blocking the Hall effect of these complexes. The second type of complexes contains a vacancy charged similar as the ion (H$^+$VD and OH$^-$VL). This second type is responsible for only a small fraction of the electric mobility (10-20\%), but their movement is not altering the electric polarisation of the lattice structure. They are responsible for the measured small Hall mobilities of the ions. 
}

\vspace{10pt}
\noindent\rule{\textwidth}{1pt}
\tableofcontents\thispagestyle{fancy}
\noindent\rule{\textwidth}{1pt}
\vspace{10pt}

\section{Introduction}
\label{sec:intro}
The electrical properties of pure water are extensively studied. Most models are based on the ideas of Svante Arrhenius (1884) explaining the electric current in water by
the movement of both H$^+$ and OH$^-$ ions due to an applied electric field~\cite{RN5677}. However, the H$^+$ and OH$^-$ mobilities were found to be a factor of 5 to 10 larger than most other ions~\cite{RN6909}. This anomaly is generally explained  by the Grotthuss mechanism~\cite{RN11914}). Not the ions themselves are moving through the lattice, but the charge of the H$^+$ and OH$^-$ions (i.e. the proton) hops from water molecule to water molecule. However, this mechanism fails to explain the positive temperature-conductivity  relation of pure water and was replaced by N. Agmon~\cite{RN6909, RN10836} to protonic hopping within H$_5$O$_2$$^+$ and H$_9$O$_4$$^+$ clusters for the H$^+$ ion and within H$_7$O$_4^-$ and (HOHOH)$^-$ clusters  for OH$^-$ ions. These clusters have to move throughout the water structure making from the cleavage of a hydrogen-bond within the clusters the rate-limiting process of the ionic mobility. This cluster model gives the mobility of H$^+$ and OH$^-$ its positive temperature dependence. 

In part I of this paper the electric behaviour of pure water is described by the Protonic-Semiconductor Interstitial-Ice model (or $\psi$ model). Liquid water consists of an intact ice-like lattice with a significant percentage of both vacant lattice positions and water molecules filling the interstitial sites of the open ice-like lattice. DC electrical conduction is mainly determined by ionic transport throughout the lattice. To explain the positive temperature dependency of the conductivity  it was assumed that each ion in water forms a complex with a vacancy, favouring a thermally-induced ionic hopping mechanism in the water lattice. This mechanism allows to explain the significant jump in the DC conductivity (a factor of 39) after the phase transition from ice to water because in ice a majority of the ions complex with DL defects blocking their motion (see part I). The clustered-ions mechanism of N. Agmon fails to explain this jump. Because the conduction mechanism is completely new and essential for the further study of electrolytes, we searched for extra evidence in favour of this new mechanism. 

The Hall effect is the occurrence of a small electrical voltage across a current-carrying electrical conductor when this conductor is placed in a transverse magnetic field~\cite{RN6852}. This effect is commonly used in semiconductor physics to characterise the semiconductor and to determine the charge density of electrons and holes, but it can
be measured in any conductor. In the 1980-90 a research group of the University of Pierre and Marie Curie succeeded to measure the small
Hall effect in dilute aqueous electrolyte solutions~\cite{RN6358,RN6318,RN6353,RN6352,RN6361}. Their findings were inconsistent with the mainstream ionic drift model unless
they assumed that the ionic mobility in the Hall field was significantly smaller from the ionic mobility in the applied electric field. They concluded that, although the mobility of H$^+$ and OH$^-$ is anomalously larger than other ions, their Hall mobility was anomalously smaller than the other ions. Several models were proposed to explain the small Hall mobilities, such as
the model of Hubbard-Wolynes~\cite{RN6854}, of Kroh-Felderhof~\cite{RN6870,RN6869} and Sung-Friedman~\cite{RN6868}, but these models fail to explain the  behaviour of the  H$^+$ and OH$^-$ ions. 
 
Within part II, the proposed DC conduction mechanism based on the complex vacancy-ion will be further explored. The goal is to find an explanation for the Hall data that is consistent with the $\psi$ model.

\section{The Hall data measured in electrolytes }
Because the conductivity of water is too small to measure accurately, the behaviour of the H$^+$ and OH$^-$ions is studied in dilute aqueous electrolytes, like HCl of NaOH solutions. 
Over a period of more than 20 years, the Hall effect of monovalent electrolytes was investigated by P. and R.~Gerard, M.~Meton, E. J.~Picard and M.~Abbes and a consisted dataset was obtained~\cite{RN6358,RN6318,RN6353,RN6352}. The methodology to measure the Hall voltage is described in~\cite{RN6361}. 

The standard setup to measure the Hall effect is when a DC current is running through a rectangular specimen and a magnetic field is applied in a direction opposite to this current~\cite{RN8435}. If the DC current is enforced by an electric field in the positive x-direction (component $E_x$) and the magnetic induction (with a positive component $B$) is applied along the positive z axes, a Hall tension along the y direction is measured corresponding to the electric field $E_y$. The reason for this is the Lorentz force on the moving charges. 
This size of the Hall effect is characterised by the Hall coefficient, defined as
\begin{equation}
R_{H}=\frac{E_{y}}{J_x B}, \label{RH}
\end{equation}
with $J_x$ the electric current density in the x direction. This coefficient will first be derived within the mainstream model of H$^+$ and OH$^-$ions drifting through the water without any interactions with the VDLs and transporting a charge with size $e$.  

The Lorentz force on H$^+$ and OH$^-$ ions will push them to the negative y direction in the specimen creating a Hall electric field $E_y$.  A H$^+$ moving with an average velocity $v_{+,x}$ in the x direction will experience a netto Lorentz force $F_ {+,y}$ equal to
\begin{equation}
F_ {+,y}=  e  E_y - e  v_{+,x} B, \label{eq:Lorentz10}
\end{equation} 
 This Lorentz force is the driving force for the electric current $J_{+,y}$ of the H$^+$ ions in the y direction. This leads to the following equations for the electric current densities of the ions,  
\begin{eqnarray}
J_{+,y} &=&  \sigma_{+} (E_y -v_{ +,x} B),  \label{eq:J10} \\
                                                                  &=&  \sigma_{+} ( E_y-\mu_{+} E_x B), \label{eq:J11}
 \end{eqnarray} 
with $\mu_{+}$ the mobility of the $H^+$ ions.  

Similar equations can be found for the OH$-$ ions moving in the negative x direction, so with a negative value for $v_{-,x}$. The Lorentz force will be
\begin{equation}
F_ {-,y}=  -e  E_y + e  v_{-,x} B, \label{eq:Lorentz11}
\end{equation} 
and the corresponding current density 
\begin{eqnarray}
J_{-,y} &=&  \sigma_{-} (E_y -v_{ -,x} B),  \label{eq:J12}, \\
                           &=&  \sigma_{-} ( E_y+\mu_{-} E_x B). \label{eq:J13}
 \end{eqnarray} 

Taking into account that the netto current density along the y axis ($J_{+,y} + J_{-,y}$) is zero, the summing Eq.~\ref{eq:J11} and Eq.~\ref{eq:J13} leads to the following relation
\begin{equation}
(\sigma_+ + \sigma_-)  E_y  = (\sigma_{+} \mu_{+} -  \sigma_{-} \mu_{-})E_x B .  \label{eq:J10}
\end{equation} 
Because $J_x = (\sigma_+ + \sigma_-) E_x$, the Hall coefficient can be found 
 \begin{equation}
R_{H}=\frac{E_{y}}{JB}= \frac{\sigma_{+} \mu_{+} -  \sigma_{-} \mu_{-}}{(\sigma_{+} +  \sigma_{-})^2 }, \label{eq:RH3}
\end{equation}

The mobilities and conductivities of the ions are accurately known, so Eq.~\ref{eq:RH3} can easily be used as a test for the experimental Hall data. However, the researchers found significant discrepancies between the measured and calculated values for all tested salts and acids~\cite{RN6358,RN6318,RN6353,RN6352}. To overcome this problem, it was assumed that the two different components of the Lorentz force have a different mobility~\cite{RN6352}. The classical electrical mobility $\mu$ which relates the drift speed $\mathbf{v}$ of the charges to the applied electric field $\mathbf{E}$
\begin{equation}
\mathbf{v=\mu\mathbf{E}},
\end{equation}
and a Hall mobility $\mu^{h}$ which relates the drift speed $\mathbf{v}$ of the charges to the Hall field $\mathbf{v\times\mathbf{B}}$
\begin{equation}
\mathbf{v}=\mu^{h}\mathbf{v\times\mathbf{B}}.
\end{equation}
Eq.~\ref{eq:RH3} was based on the assumption that both mobilities are equal. However when both mobilities are different the Hall coefficient becomes~\cite{RN6352,RN6358}
\begin{equation}
R_{H}=\frac{\sigma_{+} \mu_{+}^h -  \sigma_{-} \mu_{-}^h}{(\sigma_{+} +  \sigma_{-})^2 }.\label{eq:RH4}
\end{equation}
This equation was conform the experimental results and used to derive the Hall mobilities. To compare the effect of different ions more easily, the Hall coefficient is rewritten as $h = e n_+ R_H$ with $h$ a dimensionless parameter, also called the Hall number of the solution~\cite{RN6352}, 
\begin{equation}
h=en_{+}R_{H}=\frac{\sigma_{+}^{2}h_{+}-\sigma_{-}^{2}h_{-}}{(\sigma_{+}+\sigma_{-})^{2}},\label{eq:RH5}
\end{equation}
 and $h_{+}$ and $h_{-}$
the Hall numbers of respectively the positive and negative ions.
The Hall numbers of the individual ions are defined as the ratio of
the Hall mobility to the electric mobility of the ion, i.e.
\begin{equation}
h_{-}=\frac{\mu_{-}^{h}}{\mu_{-}}
\end{equation}
and

\begin{equation}
h_{+}=\frac{\mu_{+}^{h}}{\mu_{+}}.
\end{equation}

The experimental results are structured as Hall numbers for the individual
ions~\cite{RN6318,RN6353,RN6352,RN6361}. Some Hall numbers for different ions are found
in Table~\ref{tab:Hall-numbers}. The Hall numbers are smaller than
one, indicating that the mobility in the Hall field is smaller that
the electrical mobility. More important, the Hall numbers of H$^+$
and OH$^{-}$ have a divergent behaviour. While
the electrical mobility of both ions is anomalously high (mainstream explained by the Grotthuss mechanism) their
magnetic mobility is anomalously reduced compared with the other ions.

Several models were proposed to explain the smaller Hall mobility, such as
the model of Hubbard-Wolynes~\cite{RN6854}, of Kroh-Felderhof~\cite{RN6870,RN6869},
and Sung-Friedman~\cite{RN6868}. They make an estimate of the Hall number
between 0.71-0.75 but none of them gives a good clue for the difference
between the Hall numbers of negative and positive ions, nor for the small Hall numbers of H\protect\textsuperscript{+} and OH\protect\textsuperscript{-}~\cite{RN6352}.  

\begin{table}
\begin{centering}
\begin{tabular}{|c||c|c|c|c|}
\hline 
ion & $h_{+}$ & $\mu_{+}$ (cm\textsuperscript{2}/Vs) \tabularnewline
\hline 
\hline 
$\mathrm{H^{+}}$ & 0.1 & 0.003629 \tabularnewline
\hline 
$\mathrm{Na^{+}}$ & 0.7 & 0.000520 \tabularnewline
\hline 
$\mathrm{Li^{+}}$ & 0.8-1 & 0.000401 \tabularnewline
\hline 
$\mathrm{K^{+}}$ & 0.6-0.7 & 0.000764 \tabularnewline
\hline 
\hline 
ion & $h_{-}$ & $\mu_{-}$ (cm\textsuperscript{2}/Vs) \tabularnewline
\hline 
\hline 
$\mathrm{OH^{-}}$ & 0.2 & 0.002067 \tabularnewline
\hline 
$\mathrm{Cl^{-}}$ & 0.5 & 0.000793 \tabularnewline
\hline 
$\mathrm{I^{-}}$ & 0.5 & 0.000797 \tabularnewline
\hline 
\end{tabular}
\par\end{centering}
\caption{Hall numbers and mobilities of different ions at room temperature. For
some ions Hall numbers differ depending on the source~\cite{RN6318,RN6352}. The ionic mobilities are calculated from the molar conductivities~\cite{RN5678}. \label{tab:Hall-numbers}}
\end{table}

\section{Polarising electric currents and the Hall effect}
The electric current $\mathbf{ J} $ in an ice-like lattice in water is derived by Jaccard's theory and is based on the assumption that moving ions are electrically polarising the lattice structure in an opposite direction as the moving VDL defects. Because drift currents are opposed by diffusive currents, a single defect cannot carry a DC current and these currents will disappear with the Debye relaxation time~\cite{JDP1}. Only the combination of ionic and charged vacancies allows a stable DC current. In~\cite{JDP1} we reformulated Jaccard's theory in order to make it consisted with the macroscopic Maxwell equations. Therefore, the difference between bound and free charges must be taken into account. VDLs are transporting a bound charge $\pm e_{\mathrm{\it{DL}}} = \pm 0.38 e$ throughout the lattice, while the ions are transporting both a free charge $\pm e$ together with an opposite bound charge $\mp e_{\mathrm{\it{DL}}}$. The netto size of the charge of the ions is therefore only  $e_{\mathrm{\it{\pm}}} = 0.62 e = e -e_{\mathrm{\it{DL}}}$. 

The total electric current $\mathbf{ J}$ inside the lattice is the sum of the current of bound charges $\mathbf{ J_b}$ and the current of the free charges $\mathbf{ J_f}$ 
\begin{equation}
\mathbf{ J} =  \mathbf{ J_b} +  \mathbf{ J_f} = e_{\mathrm{\it{DL}}} (-\mathbf{j_{+}}+\mathbf{j_{-}}+\mathbf{j_{VD}} -\mathbf{j_{VL}}) + e(\mathbf{j_{+}}-\mathbf{j_{-}}),  \label{eq:J00}
\end{equation}
with  and $\bf{j_{+}}$, $\bf{j_{-}}$,  $\bf{j_{VD}}$, $\bf{j_{VL}}$ the flux density of the H$^+$ defects, the OH$^-$ defects, the VD defects and the VL defects respectively. At DC the lattice structure becomes polarised, reducing $\mathbf{ J_b}$ to 0~\cite{JDP1} and resulting in a DC current $\mathbf{ J_o}$,
\begin{equation}
\mathbf{ J_{o}} =  \mathbf{ J_f} =  e(\mathbf{j_{+}}-\mathbf{j_{-}}).  \label{eq:J02}
\end{equation}
The DC current can be seen as the movement of the free charge $e$ by the ions. However, this not what is physically happening. It is not correct to say the there is no VDL current present in DC because a single type of defect cannot sustain a DC current. Besides the free charge, the ions are also transporting an opposite bound charge with them~\cite{JDP1}. In order to balance the current of the bound charges to zero, a VDLs current $\bf{j_{VD}}$, $\bf{j_{VL}}$, is also present. A more physical way to write the same DC current is 
\begin{equation}
\mathbf{ J_{o}} =   e_\pm (\mathbf{j_{+}}-\mathbf{j_{-}}) + e_{\mathrm{\it{DL}}}  (\mathbf{j_{VD}} -\mathbf{j_{VL}}),  \label{eq:J03}
\end{equation}
with the first term the contribution of ions to the total current and the second term the contribution of the charged vacancies. Both contributions can be worked out using the current densities derived in part 1 (Eq. 69-72) and filling in $\mathbf{ P_{so}}$ as derived in~\cite{JDP1}, 
\begin{eqnarray}
\mathbf{ J_{o}} &=&  \frac{e}{e_\pm} \sigma_\pm \mathbf{ E} +  \frac{e e_{\mathrm{\it{DL}}} }{e_\pm^2} \sigma_\pm  \mathbf{ E}, \label{eq:J04} \\
                      &=& (\frac{e}{e_\pm})^2 \sigma_\pm \mathbf{ E} \label{eq:J04b},
\end{eqnarray} 
with $\sigma_\pm = \sigma_+ + \sigma_-$ the total ionic conductivity of the ions and $\mathbf{ E}$ the applied electric field. Notice that sum of both contributions in Eq.~\ref{eq:J04b} corresponds to the total DC conductivity as was derived in part 1 (Eq.~43).  

Both the ionic current (first term of Eq.~\ref{eq:J04}) and the VDL current (second term of Eq.~\ref{eq:J04}) have the same order of magnitude. Although the concentration of VDLs is several orders of magnitudes larger than the ionic concentration, the netto VDL currents are a factor of 2 smaller than the netto ionic currents. This will have its impact on the mobilities of both the ions and the VDLs. From the first term of Eq.~\ref{eq:J04} one can extract the total ionic conductivity $\sigma_\pm^t$   
\begin{equation}
\sigma_\pm^t =  e_\pm (n_+  \frac{e\mu_+}{e_\pm} + n_- \frac{e\mu_+}{e_\pm}),   \label{eq:J05}
\end{equation} 
showing that the total mobility of the ions is a factor $\frac{e}{e_\pm}$ larger than the drift mobility. Similar the total conductivity of the VDLs can be derived out of the second term of Eq.~\ref{eq:J04},
 \begin{equation}
\sigma_{\mathrm{\it{VDL}}}^t = e_{\mathrm{\it{DL}}} (n_+  \frac{e \mu_+ }{e_\pm} + n_- \frac{e\mu_+}{e_\pm}).   \label{eq:J06}
\end{equation} 
This conductivity is a result of all the movement of all the VDLs present in the lattice, so the total mobility of both the VDs and the VLs are 
\begin{eqnarray}
 \mu_{\mathrm{\it{VD}}}^t &= & \frac{en_+\mu_+}{e_\pm n_VD},    \label{eq:J07} \\
 \mu_{\mathrm{\it{VL}}}^t &= & \frac{en_-\mu_+}{e_\pm n_VL},    \label{eq:J08}  
\end{eqnarray} 
showing that the mobilities (and the corresponding average velocities) of the VDLs are around 7 orders of magnitude smaller than the mobility (or the average velocity) of the ions. This makes clear that the magnetic component of the Lorentz force will be negligible on the VDLs. A Hall electric field can only be formed by the movement of ions downwards. However, ions in a lattice can only move in the y direction if they are compensated by VDLs. Because there is no significant driving force on the VDLs, the ions aren't coupled with to the ionic movement and no Hall effect will occur. This absence can also be explained using quantum mechanical models~\cite{RN975, RN794}, but our semi-classical explanation links the problem more clearly to the presence of polarising currents in the lattice.  

Experiments show that in ice no Hall effect is found~\cite{RN975, RN794}. In water the situation is more complex. A Hall effect is found for  H$^+$ and OH$^-$ ions but it is significantly smaller than the Hall effect of the other ions~\cite{RN6358,RN6318,RN6353,RN6352}. Probably is the initial assumption, the fact that all moving ions are polarising the lattice, is not completely correct in water. 

\section{Ionic conduction mechanisms in water}
As was already proven in part I of this paper, the Grotthuss mechanism is not responsible for the DC conductivity in water. Instead, we found that a hopping mechanism of the molecules, where ions interact with a vacancy, is much more consistent with the conductivity data that has a positive temperature dependency. In this section we will discuss the details of this mechanism and show that dependent of the type of vacancies involved, the effect of the moving ions on the lattice polarisation will differ. 

\begin{figure*}
\centering{\includegraphics[width=100mm]{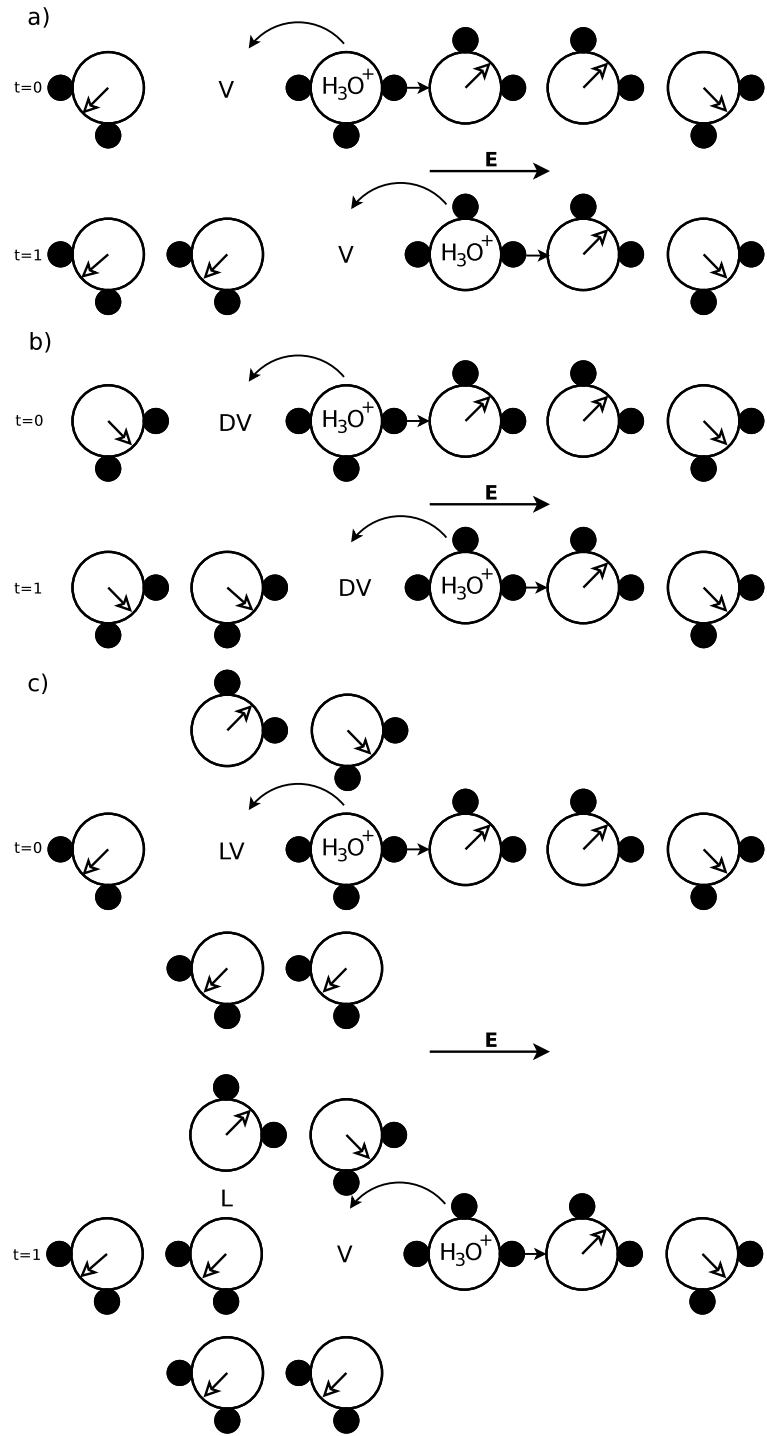}}
\caption{Schematic 2D representation of the movement of three types of ionic defects under influence of an external electric field $\bf{E}$: a) a H$^+$V complex, b) a H$^+$VD complex c) a H$^+$VD complex. The first row is the starting chain, the second row is the water chain after the jumping of the complex. We indicated the dipole moment of the individual water molecules $\mathbf{p_o}$ with the small arrows. Notice the change in polarisation density after the movement of the complex in a en c and the fact that this not happen in b. }
\label{fig:2Dmodel1}
\end{figure*}

In Fig. 9 of part I of this paper, a first suggestion was made how a VL allows a H$^+$ ion to jump through the lattice. However, the movement was not in the direction of the electric field (the x direction). In order to move easily through the lattice, the H$^+$ ion has to from a complex with the vacancies. There are three types of H$^+$V complexes possible, and all three allow the ions to move several steps in the x direction. The three possibilities (a, b and c) are drawn in Fig.~\ref{fig:2Dmodel1}. Notice that the vacancy is situated behind the ion and that in order to do several jumps in the positive x direction two mechanisms have to be combined: the tunnelling of the proton to the next water molecule together with a jump of the water molecule. In case a, the H$^+$V complex, with a neutral vacancy, can move through the lattice, thereby changing the polarisation of the lattice. In case b, the H$^+$VD complex can move through the lattice but now, the lattice polarisation isn't changed. In case c, we found only one way  to combine a VL with H$^+$ which allows the travel through the lattice. Notice that LV is split up in a vacancy and an L defect. The ion can move further as an  H$^+$V complex, but is also likely that the L defect will be attracted to the positively charged  H$^+$V complex to form again an H$^+$VL. This mechanism is also changing the polarisation of the lattice. 

So, three mechanisms are available for the movement of H$^+$. Taking into account the positive charge of the H$^+$ ion, the third mechanism with the H$^+$VL complex, is the most likely to occur. Much lesser likely are H$^+$VD complexes because of the repulsive force between them.  A similar figure can be designed for the OH$^-$ ions. In this case the movement of OH$^-$V and  OH$^-$VD complexes will polarise the lattice and the OH$^-$VL will leave the lattice as it was. 

Because the H$^+$VD and OH$^-$VL complexes are not altering the lattice polarisation, they can move through the lattice without any extra compensation. So, they are not part of Jaccard's theory and they obey the normal conduction rules. So the total DC conductivity in water can be written as
\begin{eqnarray}
\sigma_{o} &=&  (\frac{e}{e_\pm})^2 (\sigma_{\scriptscriptstyle  \mathrm{  H^+\!V}} + \sigma_{\scriptscriptstyle \mathrm{H^+\!VL}})  +  \sigma_{\scriptscriptstyle \mathrm{H^+\!VD}}  + (\frac{e}{e_\pm})^2 (\sigma_{\scriptscriptstyle \mathrm{OH^-\!V}} + \sigma_{\scriptscriptstyle \mathrm{OH^-\!VD}} )  + \sigma_{\scriptscriptstyle \mathrm{OH^-\!VL}} ,  \label{eq:sigmao} \\ 
                  &=& \sigma_{+} + \sigma_{-}.  \label{eq:sigmao2} 
\end{eqnarray} 
with the index referring to the type of complex and $\sigma_+$ and $\sigma_-$ the contribution of the positive and negative ions to the conductivity, respectively.  


\section{Non-polarising electric currents and the Hall effect }
The non-polarising currents of the complexes H$^+$VD and OH$^-$VL are not compensated by VDL defects. Therefore, these complexes are sensitive to the Hall effect. In this section, an equation for the Hall coefficient $R_H$ for these complexes wil be derived. 

The Lorentz force on H$^+$VD and OH$^-$VL will push them to the negative y direction in the specimen creating a Hall electric field $E_y$. Notice that the magnitude of the total charge of these complexes is $e$ and not $e_\pm$. A moving H$^+$VD will experience a netto Lorentz force $F_ {\mathrm{HVD,y}}$ equal to
\begin{equation}
F_ {\scriptscriptstyle \mathrm{H^+\!VD,y}}=  e  E_y - e  v_{\scriptscriptstyle \mathrm{H^+\!VD,x}} B, \label{eq:Lorentz1}
\end{equation} 
with $v_{\scriptscriptstyle \mathrm{H^+\!VD,x}}$  the average velocity of H$^+$VD in the x direction. This Lorentz force is the driving force for the electric current $J_{\mathrm{H^+\!VD,y}}$ of the H$^+$VD in the y direction. When $\mu_{\scriptscriptstyle \mathrm{H^+\!VD}}$ is the mobility of the H$^+$VD complexes this gives 
\begin{eqnarray}
J_{\scriptscriptstyle \mathrm{H^+\!VD,y}} &=&  \sigma_{\scriptscriptstyle \mathrm{H^+\!VD}} (E_y -v_{\scriptscriptstyle \mathrm{H^+\!VD,x}} B),  \label{eq:J1} \\
                                                                  &=&  \sigma_{\scriptscriptstyle \mathrm{H^+\!VD}} ( E_y-\mu_{\scriptscriptstyle \mathrm{H^+\!VD}} E_x B). \label{eq:J2}
 \end{eqnarray} 
Similar equations can be derived for OH$^-$VL complexes, now moving in the negative x direction, so with a negative value for $v_{\mathrm{OHVL,x}}$. The Lorentz force will be
\begin{equation}
F_ {\scriptscriptstyle \mathrm{OH^-\!VL,y}}=  -e  E_y + e  v_{\scriptscriptstyle \mathrm{OH^-\!VL,x}} B, \label{eq:Lorentz2}
\end{equation} 
and with $\mu_{\scriptscriptstyle \mathrm{OH^-\!VL}}$ the mobility of the OH$^-$VL  complexes, 
\begin{eqnarray}
J_{\scriptscriptstyle \mathrm{OH^-\!VL,y}}   &=&  \sigma_{\scriptscriptstyle \mathrm{OH^-\!VL}} (E_y -v_{\scriptscriptstyle \mathrm{OH^-\!VL,x}} B),  \label{eq:J3} \\
                                                         &=&  \sigma_{\scriptscriptstyle \mathrm{OH^-\!VL}} (E_y +\mu_{\scriptscriptstyle \mathrm{OH^-\!VL}} E_x B). \label{eq:J4}
 \end{eqnarray} 
A Hall electric field is now able to build up. This field will also move the other complexes H$^+$V, H$^+$VL, OH$^-$V and OH$^-$VD  and their corresponding VDLs.  However, they are only driven by the electric component of the Lorentz force, not by the magnetic field. The electric current densities of the complexes are 
 \begin{eqnarray}
 J_{\scriptscriptstyle \mathrm{H^+\!V,y}} &=& (\frac{e}{e_\pm})^2 \sigma_{\scriptscriptstyle \mathrm{H^+\!V}} E_y, \label{eq:J5}\\
J_{\scriptscriptstyle \mathrm{OH^-\!V,y}} &=&   (\frac{e}{e_\pm})^2 \sigma_{\scriptscriptstyle \mathrm{OH^-\!V}} E_y,  \label{eq:J6}\\
J_{\scriptscriptstyle \mathrm{H^+\!VL,y}} &=&   (\frac{e}{e_\pm})^2 \sigma_{\scriptscriptstyle \mathrm{H^+\!VL}}  E_y,  \label{eq:J7}\\
J_{\scriptscriptstyle \mathrm{OH^-\!VD,y}} &=&   (\frac{e}{e_\pm})^2 \sigma_{\scriptscriptstyle \mathrm{OH^-\!VD}} E_y.  \label{eq:J8}
 \end{eqnarray} 
 The netto current in y direction: $J_{\scriptscriptstyle \mathrm{H^+\!VD,y}}+ J_{\scriptscriptstyle \mathrm{OH^-\!VL,y}}+ J_{\scriptscriptstyle \mathrm{H^+\!V,y}}+ J_{\scriptscriptstyle \mathrm{OH^-\!V,y}} + J_{\scriptscriptstyle \mathrm{H^+\!VL,y}}+ J_{\scriptscriptstyle \mathrm{OH^-\!VD,y}}$ is zero. So the sum of Eqs.~\ref{eq:J2}, \ref{eq:J4}, \ref{eq:J5}-\ref{eq:J8} will also be zero, relating $E_y$ to $E_x$ and B. Applying the definition of the total conductivity (Eq.~\ref{eq:sigmao}) this relation will look like
 \begin{equation}
\sigma_o E_y   = (\sigma_{\scriptscriptstyle \mathrm{H^+\!VD}} \mu_{\scriptscriptstyle \mathrm{H^+\!VD}} -  \sigma_{\scriptscriptstyle \mathrm{OH^-\!VL}} \mu_{\scriptscriptstyle \mathrm{OH^-\!VL}})E_x B.  \label{eq:J20}
\end{equation} 
Because $J_x = \sigma_o E_x$, the Hall coefficient can be found 
\begin{equation}
R_{H}=\frac{E_{y}}{J_x B} = \frac{ \sigma_{\scriptscriptstyle \mathrm{H^+\!VD}} \mu_{\scriptscriptstyle \mathrm{H^+\!VD}} -  \sigma_{\scriptscriptstyle \mathrm{OH^-\!VL}} \mu_{\scriptscriptstyle \mathrm{OH^-\!VL}}}{\sigma_o^2}.  \label{RH10}
\end{equation} 
Within the dimensionless formulation ($h = e n_+ R_H$)~\cite{RN6352} these Hall mobilities are related to Hall numbers. Dividing $\sigma_o$ in the contribution of both the positive as the negative charges, as was done in Eq.~\ref{eq:sigmao2} and assuming that $n_+ = n_-$ one can rewrite h as   
\begin{eqnarray}
h=en_{+}R_{H}&=&\frac{ \sigma_{\scriptscriptstyle \mathrm{H^+\!VD}} e n_+ \mu_{\scriptscriptstyle \mathrm{H^+\!VD}} -  \sigma_{\scriptscriptstyle \mathrm{OH^-\!VL}} e n_-\mu_{\scriptscriptstyle \mathrm{OH^-\!VL}}}{(\sigma_o^+ + \sigma_o^-)^2} ,  \label{eq:RH11}\\
          &=&\frac{ \sigma_+^2 \frac{\sigma_{\scriptscriptstyle \mathrm{H^+\!VD}} e n_+ \mu_{\scriptscriptstyle \mathrm{H^+\!VD}}}{\sigma_+^2} -\sigma_-^2 \frac{\sigma_{\scriptscriptstyle \mathrm{OH^-\!VL}} e n_-\mu_{\scriptscriptstyle \mathrm{OH^-\!VL}}}{\sigma_-^2 } }{(\sigma_+ + \sigma_-)^2} , \label{eq:RH12}\\
                    &=&\frac{ \sigma_+^2 h_+ -\sigma_-^2 h_-}{(\sigma_+ + \sigma_-)^2}  \label{eq:RH13}.
\end{eqnarray}
Eq.~\ref{eq:RH12} and ~\ref{eq:RH13} define the Hall numbers $h_+$ and $h_-$ for both ions within the $\psi$ model. If we assume that the mobilities of the different complex types are of the same order of magnitude, and if the charge correction factors in Eq.~\ref{eq:sigmao} are ignored, the size of Hall numbers can easily be estimated as 
\begin{eqnarray}
h_+ &=& \frac{\sigma_{\scriptscriptstyle \mathrm{H^+\!VD}} e n_+ \mu_{\scriptscriptstyle \mathrm{H^+\!VD}}}{\sigma_+^2}, \label{eq:h+1}\\
                    &\approx&\frac{ n_{\scriptscriptstyle \mathrm{H^+\!VD}}}{n_+} , \label{eq:h+2}\\
 h_- &=& \frac{\sigma_{\scriptscriptstyle \mathrm{OH^-\!VL}} e n_+ \mu_{\scriptscriptstyle \mathrm{OH^-\!VL}}}{\sigma_-^2}, \label{eq:h-1}\\
                    &\approx&\frac{ n_{\scriptscriptstyle \mathrm{OH^-\!VL}}}{n_-}.  \label{eq:h-2}
\end{eqnarray}
So the Hall numbers are a first estimation of the fraction of the complexes H$^+$VD and OH$^-$VL that do not polarise the specimen. 

\section{Discussion}
The outline of this paper is pure water, so we will focus here on the Hall number of H$^+$ ions, i.e. 0.1, and the Hall number of OH$^-$ which is double as large, i.e. 0.2. This doubling is interesting, because the mobility of the OH$^-$ is around half of the mobility of the H$^+$ ion (factor 0.56). Within the $\psi$ model, the three types of vacancies are responsible of the ionic mobility by forming complexes with the ions. Because of their electrostatic attraction it is reasonable to assume that H$^+$VL and OH$^-$VD complexes are the most dominant and they will dominantly determine the mobility values of the ions. The mobility values suggest that the density of VLs is a factor of 2 larger than the density of VDs in water. This finding is conform with a similar finding in ice. In ice, the Bjerrum D defects are all trapped to interstitials. De Koning~\cite{RN10913} used a first-principles modelling to prove this in the ice lattice. Similar calculations in the water lattice, containing a high density of vacancies and interstitials, may prove the only partially trapping of D defects providing extra evidence for the $\psi$ model.

The Hall numbers in water provide extra evidence for the trapping of the D defects, diminishing the occurrence of VDs in water. As was proven in Eq.~\ref{eq:h+2} and \ref{eq:h-2} are the Hall numbers the fraction the non-polarising complexes. For the H$^+$ ions around 10\% of the ions are a H$^+$VD complex, for the OH$^-$ ions around 20\% of the ions are a OH$^-$VL defect, a result conform with the trapping hypothesis of D defects to interstitials. 

The Hall effect is also valuable for the study of ions in water. However, the behaviour of the different ions in water is more complex. Besides the Hall numbers, the conductivity, diffusion coefficient, 't Hoff factors, and the proton resonance frequency of electrolytes significantly differ depending on the sign and size of the different atoms. A study combining all this data will be published elsewhere.
 
\section{Conclusions}
The Hall effect of H$^+$ and OH$^-$ ions is explained based on the formation of ionic complexes with vacancies. Two types of complexes are found. The first type of complexes contains both vacancies of the opposite charge of the ion (H$^+$VL and OH$^-$VD) and neutral vacancies (H$^+$V and OH$^-$V). The movement of these complexes is changing the electric polarisation of the lattice structure which makes them insensitive for the Hall effect. The second type of complexes contains a vacancy charged similar as the ion (H$^+$VD and OH$^-$VL) and these moving vacancies are not altering the electric polarisation of the lattice structure. They are responsible for of only a small fraction of the electric mobility (10-20\%) but  responsible for the small Hall mobilities of the ions.

\section*{Acknowledgements} 
The author thanks Ward De Jonghe, Prof. Em. Roland Van Meirhaeghe and Jana De Poorter for their support and their critical reviews of this work. 

\begin{appendix} 
\end{appendix}



\bibliography{literatuur}

\nolinenumbers

\end{document}